\documentclass[]{aastex631}
\usepackage{xspace}
\newcommand{\roma}[1]{\uppercase\expandafter{\romannumeral#1}}
\newcommand{\speed}[1]{#1\,km\,s${}^{-1}$}

\newcommand{\nfig}[1]{Figure~\ref{#1}}

\newcommand{\alf}{Alfv\'en~}
\newcommand{\ha}{H{\ensuremath{\alpha}}\xspace}

\newcommand{\m}{Moreton\xspace}

\usepackage{amsmath}
\usepackage{subfigure}
\usepackage{hyperref}%
\usepackage{gensymb}

 \usepackage{url}
\usepackage{graphicx}
\usepackage{natbib}
\usepackage{amssymb,txfonts}
\usepackage{multirow}
\usepackage{array}
\usepackage{sidecap}
\usepackage{hyperref}
\usepackage{epstopdf}
\usepackage{footnote}
\usepackage{tabularx}
\usepackage{booktabs}

\usepackage{CJK}

\hypersetup{
 bookmarks=true, 
 unicode=false, 
 pdftoolbar=true, 
 pdfmenubar=true, 
 pdffitwindow=false, 
 pdfstartview={FitH}, 
 pdftitle={My title}, 
 pdfauthor={Author}, 
 pdfsubject={Subject}, 
 pdfcreator={Creator}, 
 pdfproducer={Producer}, 
 pdfkeywords={keyword1} {key2} {key3}, 
 pdfnewwindow=true, 
 colorlinks=true, 
 linkcolor=blue, 
 citecolor=blue, 
 filecolor=blue
 urlcolor=blue
 }

\shorttitle{3D fast-mode  Wave Propagation from Corona to Chromosphere: Triggering Mechanism for 3D Oscillations of filaments}
\shortauthors{Zhou et al.}

\begin{document}
\begin{CJK*}{UTF8}{gbsn}
\title{3D fast-mode Wave Propagation from Corona to Chromosphere: Triggering Mechanism for 3D Oscillations of filaments}

\author[0009-0001-0368-3543]{Yunxue Huang }
\altaffiliation{These authors contributed equally to this work and should be considered co-first authors.}
\affiliation{ College of Physics and Electronic Engineering, Sichuan Normal University, Chengdu 610068, People's Republic of China}

\author[0009-0002-9605-3547]{Qin Feng}
\altaffiliation{These authors contributed equally to this work and should be considered co-first authors.}
\affiliation{ College of Physics and Electronic Engineering, Sichuan Normal University, Chengdu 610068, People's Republic of China}

\author[0000-0003-2183-2095]{Yuhu Miao}
\affiliation{School of Information and Communication, Shenzhen Institute of Information Technology, Shenzhen, 518172, People's Republic of China}

\author[0000-0001-8318-8747]{Zhining Qu}
\affiliation{ College of Physics and Electronic Engineering, Sichuan Normal University, Chengdu 610068, People's Republic of China}

\author[0000-0002-1134-4023]{Ke Yu}
\affiliation{ College of Physics and Electronic Engineering, Sichuan Normal University, Chengdu 610068, People's Republic of China}

\author{Hongfei Liang}
\affiliation{Department of Physics, Yunnan Normal University, Kunming 650500, People's Republic of China}

\author{Yu Liu}
\affiliation{School of Physical Science and Technology, Southwest Jiaotong University, Chengdu, 611756, Sichuan, China}

\author[0000-0001-9374-4380]{Xinping Zhou }
\affiliation{ College of Physics and Electronic Engineering, Sichuan Normal University, Chengdu 610068, People's Republic of China}

\correspondingauthor{Yu Liu}
\email{lyu@swjtu.edu.cn}
\correspondingauthor{Xinping Zhou}
\email{xpzhou@sicnu.edu.cn}

\begin{abstract}
\m waves are widely regarded as the chromospheric counterpart of extreme ultraviolet (EUV) waves propagating in the corona. However, direct observational evidence confirming their simultaneous propagation across multiple atmospheric layers—from the corona through the transition region to the chromosphere—has been lacking. In this study, we present comprehensive observational evidence of a three-dimensional (3D) fast-mode wave propagating from the corona through the transition region into the chromosphere, exhibiting a gradual deceleration. Additionally, this wave interacts with three filaments (F1, F2, and F3) along its path, inducing oscillation with multiple amplitudes: Filaments F1 and F2 exhibit simultaneous horizontal and vertical large-scale oscillations ($\sim$\speed{20}), while Filament F3 only exhibits vertical small-scale oscillation ($\sim$\speed{4}). Interestingly, F1 displays a similar oscillation period of about 500\,s in both horizontal and vertical directions, whereas F2 shows significantly different periods in these two dimensions (1100\,s and 750\,s), and F3 exhibits only a vertical oscillation with a period of about 450\,s. Based on this kinematic behavior, we propose that their oscillations were likely triggered by compression from the flanks of the dome-shaped wavefront. We further estimate the magnetic fields of the filaments. The radial (axial) magnetic fields for F1 and F2 are estimated to be 14.9\,G (28.6\,G) and 9.9\,G (18.6\,G), respectively. For F3, we estimate its radial magnetic field to be 16.6\,G.
\end{abstract}

\keywords{Solar coronal waves(1995) --- Alfv\'en waves (23) --- Solar corona (1483)}

\section{Introduction} 
\label{se:intro}
Moreton wave, flare-associated waves seen in \ha, rapidly propagating in restricted angles with speeds range from 550-\speed{2500} \citep{1960AJ.....65U.494M,2001ApJ...560L.105W,2012ApJ...745L..18A,2013ApJ...773..166L}. At some distance ($\sim$100 Mm) from the flare site, they often appear as arc-shaped fronts, with an angular width of about 100$^\circ$, propagating away from the flare kernel which last for only minutes and soon become irregular and diffuse. As the waves were found to be too fast to originate in the chromosphere, \cite{1968SoPh....4...30U} developed a theory that interprets \m waves as the ``hemline'' of a dome-shaped magetohydrodynamic (MHD) wavefront sweeping across the chromosphere. According to this model, the increased pressure behind the wavefront compresses the chromospheric plasma, which manifests as the characteristic down-up swing in \ha filtergrams and has recently been confirmed via extreme ultraviolet (EUV) redshifts \citep{2011ApJ...737L...4H,2011ApJ...743L..10V} and direct EUV imaging \citep{2012ApJ...753...52L,2011ApJ...732L..20C}. The Extreme-ultraviolet Imaging Telescope \citep[EIT;][]{1995SoPh..162..291D} on board SOHO for the first time imaged such wave-like disturbance in solar corona \citep{1998GeoRL..25.2465T}. Given that the \alf speed increases with altitude in the lower corona, the wavefront of a coronal fast-mode MHD wave is anticipated to tilt forward toward the solar surface. This inclination has been confirmed in several limb events, as documented by many studies \cite[such as][]{2011ApJ...738..160M,2012ApJ...745L...5C,2012ApJ...753...52L,2020ApJ...894...30W,2021SoPh..296..169Z}. In certain numerical simulations, the Moreton waves have been successfully reproduced. For instance, the hybrid model developed by \cite{2011ApJ...732L..20C} demonstrates how piston-driven shocks, generated by expanding coronal mass ejections (CMEs), propagate through the chromosphere and produce the \m wave signatures observed in \ha emissions. However, no comprehensive images evidence simultaneously covering corona-transition region-chromosphere have been reported to date. It is worth noting that many of the coronal EUV wave studies revealed a close association with the CMEs \citep{2005ApJ...622.1202C,2006ApJ...641L.153C,2009ApJ...698L.112C}. However, some observation and simulations reveal that the energy release during the flare can drive the EUV wave train \citep[e.g.,][]{2021ApJ...911L...8W,2022ApJ...941...59Z,2022ApJ...930L...5Z,2024ApJ...968...85Z}. Furthermore, \cite{2024SCPMA..6759611Z,2024ApJ...974L...3Z} find that the EUV wave have close relationship with the untwisting of the erupting filament. These discrepancies indicate that we need to continue research on the origin of the EUV waves.

The EUV wave always interacts with the coronal magnetic structures, such as coronal holes \citep{2012ApJ...746...13L,2022A&A...659A.164Z,2024NatCo..15.3281Z,2025ApJ...987L...3Z}, active regions \citep{2019ApJ...871L...2M}, filaments (or prominences) \citep{2012ApJ...760L..10L,2013ApJ...773..166L}, coronal loops \citep{2013Ap&SS.345...25S,2013ApJ...777...17S,2022ApJ...937L..21Z} and  coronal cavity \citep{2018ApJ...860..113Z,2023A&A...673A.154L,2023ApJ...953L..13L}. Following this interaction, filaments (or prominences) are always observed to exhibit oscillations. According to the oscillation direction, the filament oscillations can be divided into three categories, namely longitudinal oscillation \citep[e.g.,][]{2012ApJ...750L...1L,2012A&A...542A..52Z,2013A&A...554A.124Z,2014ApJ...785...79L,2014ApJ...790..100B,2014ApJ...795..130S,2017ApJ...842...27Z,2020A&A...635A.132Z,2021A&A...654A.145L,2024A&A...691A.354L,2025ApJ...981..139Y}, transverse horizontal \citep[e.g.,][]{1969SoPh....6...72K,2012ApJ...761..103G,2020A&A...642A.159Z,2024MNRAS.533.3255Z} and transverse vertical \citep{1966ZA.....63...78H,2014ApJ...786..151S}. In recent years, the simultaneous observations of two or more mixed oscillation modes have also attracted attention, like simultaneous transverse and longitudinal oscillations \citep{2017ApJ...851...47Z,2020A&A...633A..12M,2022MNRAS.516L..12T,2023MNRAS.520.3080T}, or simultaneous horizontal and vertical oscillations \citep{2023ApJ...959...71D,2025PhFl...37c7177O}. \textcolor{black}{{When the filaments are not at the solar limb, their oscillation will exhibit an interesting phenomenon in the line-of-sight (LOS) direction, namely the ``winking'' phenomenon, which is characterized by their gradual fading or complete disappearance followed by subsequent reappearance in \ha line wings. Traditionally, ``winking'' oscillations have been considered vertical \citep{2009SSRv..149..283T}. However, this characterization is accurate only when the filament is located near the disk center. For filaments closer to the limb, the oscillations may have a significant horizontal component. In this work, we use the term vertical for simplicity, although we actually refer to motions along the line of sight.}}
Notably, the observed oscillation periods primarily correlate with the filaments' intrinsic characteristic periods, showing no significant correlation with the physical dimensions of the filaments, nor with the proximity and energy magnitude of the triggering flare events. \cite{1966AJ.....71..197R} and \cite{1966ZA.....63...78H} studied 11 winking filaments, deriving oscillatory periods ranging from 6 to 40 minutes and damping times between 7 and 120 minutes. They suggested that each filament has its own characteristic frequency of oscillation, noting that a filament perturbed by waves from four flares over three consecutive days oscillated with essentially the same frequency and damping time, further emphasizing the lack of correlation between the oscillation frequency and the filament dimensions, the distance to the perturbing flare, or its size. This finding was further confirmed by \cite{2011A&A...531A..53H}, who proposed that large-amplitude filament oscillations are actually a collection of separate but interacting fine threads. Oscillations triggered by \m or coronal waves often exhibit large amplitudes \citep[$>$\speed{20}; see the review by][]{2009SSRv..149..283T}, in contrast to the commonly observed small-amplitude oscillations \citep[$\sim$2-\speed{3}; see the review by][]{2018LRSP...15....3A}, which are typically local and appear intrinsic to the filament itself, lacking any obvious external trigger. The physical mechanism underlying wave-filament interactions remains poorly understood due to the relative scarcity of observational data. Notably, filaments do not always exhibit oscillatory behavior when waves pass through them \citep{2004ApJ...608.1124O}.

Despite the fact that EUV waves have been frequently observed in the solar corona for over a decade, their counterparts in the chromosphere are rarely detected \citep{2023ApJ...949L...8Z}. Moreover, comprehensive observational evidence of their propagation through various atmospheric layers is still lacking. According to the statistic 640 coronal EUV waves from 2010 to 2021 by \cite{2025ApJ...980..254W}, they find that only 92 events (accounting for $\sim$$14.4\%$) exhibit responses in the transition region, and among these 92 events, only 20 (accounting for $\sim$$3\%$) showed chromospheric response. Typically, in wave-filament interactions, the material undergoes vertical oscillations with respect to the solar surface. In addition, simultaneous horizontal and longitudinal oscillations of a filament have been reported. However, the simultaneous horizontal and vertical (winking phenomenon) oscillations in  filaments were rarely reported in previous studies. In this paper, we present observational evidence of a coronal wave propagating from the corona to the chromosphere, and causing the filaments' three-dimensional oscillations. Section \ref{se:Observations} briefly introduces the observation instruments. Section \ref{se:results} analyzes the propagation of the wave and the oscillations of the filaments. Section \ref{se:discussion} presents the discussion and conclusions.

\section{Observations}
\label{se:Observations}
An intense flare, GOES X5.8, occurred in the active region NOAA 13664 on 11 May 2024. The flare started at 01:10 UT, peaked at 01:23 UT and ended at 01:39 UT. We observed a coronal EUV wave, accompanied by a Moreton wave, associated with the flare in the Atmosphere Imaging Assembly \citep[AIA;][]{2012SoPh..275...17L} EUV images and \ha images obtained from the Solar Magnetic Activity Research Telescope \citep[SMART;][]{2004SPIE.5492..958U}. The wave triggers multiple mini filament oscillations both in the EUV and \ha images. For analyzing the EUV wave and horizontal filament oscillations, we utilized the high spatio-temporal resolution observational data from AIA on board the Solar Dynamics Observatory \citep[SDO;][]{2012SoPh..275....3P}. The AIA takes full-disk images at seven EUV wavebands that can observe the sun from chromosphere to corona. It is well known that both the Chinese \ha Solar Explorer \citep[CHASE;][]{2019RAA....19..165L,2022SCPMA..6589602L} and SMART \citep{2004SPIE.5492..958U,2013PASJ...65...39I} instruments are capable of providing full-disk Doppler images. Unfortunately, CHASE was not operating during the period of this eruption event. Therefore, we mainly relied on data obtained from SMART to study the \m wave and the vertical oscillation of filaments using the \texttt{Cloud Model} \citep{1964PhDT........83B}. SMART regularly provides the full-disk Sun in seven wavelengths around the \ha line (6562.8\,\AA), i.e., \ha center and six off-band ($\pm0.5$\,\AA, $\pm$0.8\,\AA\, and $\pm$1.2\,\AA) with a time cadence 2 minutes and pixel size $0\arcsec.56$. Such full-disk and multi-wavelength observation with high cadence is suit to detect \m waves in the chromosphere. SUTRI utilizes the 465\,\AA\ waveband to conduct full-disk dynamic imaging observation of the solar transition region (the layer between the solar chromosphere and corona) with a spatial and temporal resolutions of about $8.\arcsec0$ and 30\,s, respectively \citep{2017RAA....17..110T,2023RAA....23f5014B}, thereby establishing a crucial bridge between the lower and upper solar atmosphere for observing the EUV wave.

\section{Results}
\label{se:results}
\subsection{The EUV wave and Moreton wave}
\begin{figure}
	\centering
	\includegraphics[width=.75\linewidth]{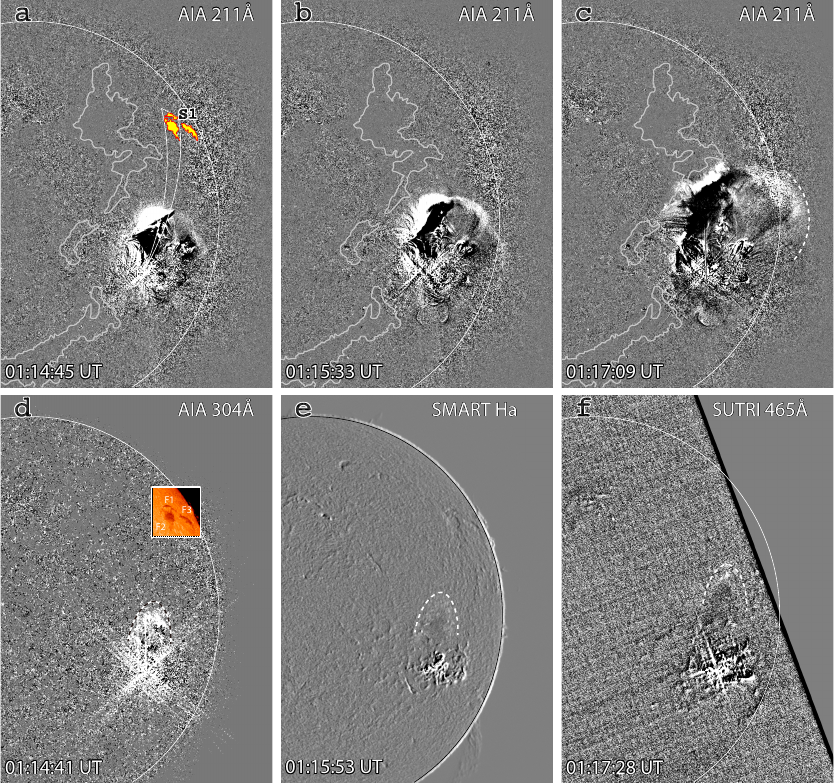}
	\caption{Overview of the EUV wave and Moreton wave. Panels (a)-(d) are the running difference of the AIA 211\, \AA\ and 304\,\AA\ images showing the EUV wave. Panel (e) exhibits the Morton wave in SMART \ha image, while panel (f) displays the EUV wave at SUTRI 465\,\AA\ running difference image. The wave front in AIA 304\, \AA, SUTRI 465\,\AA, and SMART \ha image is highlighted with white arrows. The color box in panel (d) display the mini filaments observed by CHASE \ha. The closed curves in Panels (a)-(c) marked the location of the CHs, while the sector (labeled with S1) in panel (a) is used to obtain the time-distance stack plots in Figure \ref{fig:tdp}. An online $\sim$2\,s animation  which includes 335\,\AA, 304\,\AA, 211\,\AA, 171\,\AA, 131\, \AA, 094\,\AA, SUTRI 465\,\AA\ and SMART \ha running-difference images covering 01:13 UT to 01:25 UT is available to view the details about the EUV wave and \m wave.
\label{fig:overview}}
\end{figure}

\nfig{fig:overview} shows the wave evolution observed in the corona by AIA 211\,\AA\ ((a)-(c)), transition region by SUTRI 465\,\AA\ and chromosphere by AIA 304\,\AA\ and SMART \ha center images, using the running difference images. The wave was seen from about 01:13 UT to 01:26 UT with all AIA images and propagated forward to the northeast, during this 13 minutes. Comparing the spatial structure of the wave in different bands, we can find that the \m wave (marked with a white dotted curve in \nfig{fig:overview} (e)), observed in \ha image, is well coincident with the sharp bright wavefront captured in EUV bands, during the initial stage (see panels (a), (d), (e) and (f) in \nfig{fig:overview}). It is worth mentioning the shape of the \m wave in chromosphere is slightly smaller than that of the EUV wave in the corona, which was restricted to propagate within a smaller angular range, as shown in panel (a), (e) and (f) of \nfig{fig:overview}. Additionally, we observed an expanding dome (marked with a white dotted curve in \nfig{fig:overview} (c)) that smoothly merged with the sharp one. This indicates that the EUV should be in a dome shape. These observation features support the theory proposed by \cite{1968SoPh....4...30U} that the expanding dome is thought to be the shock, the \m wave and the sharp EUV wave are the tracks when it intersection with the chromosphere and solar corona, respectively. As shown in \nfig{fig:overview} (a)-(c), three coronal holes (CHs), marked with gray curves, were scattered around the eruption origin, and three filaments (F1, F2 and F3) (see the box in panel (d) observed by CHASE \ha) on the path of the wave propagation. Please refer to the animation of \nfig{fig:overview} for the details about the EUV wave and \m wave.

\begin{figure}
	\centering

	\includegraphics[width=.65\linewidth]{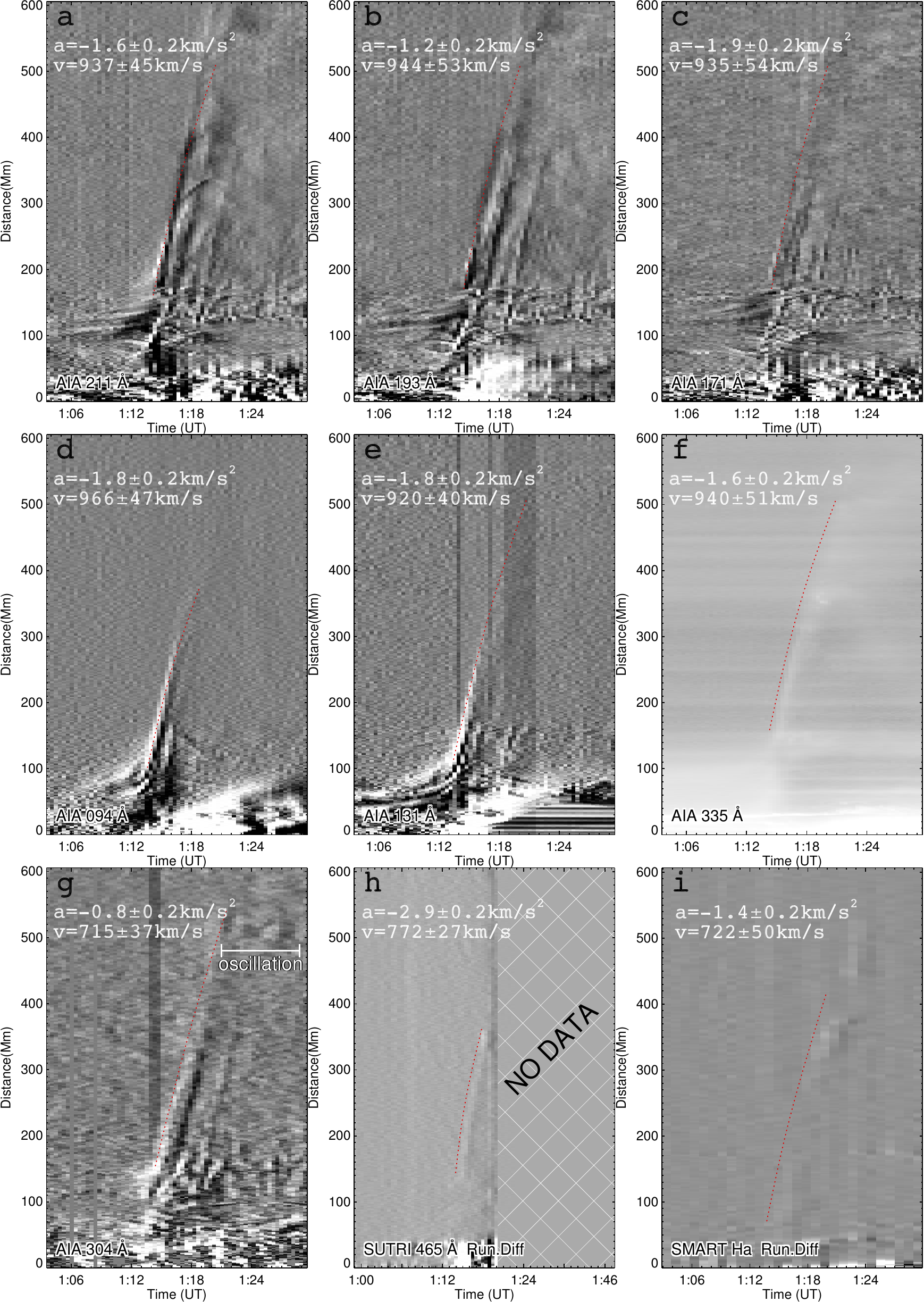}
	\caption{Time-distance stack plots showing the evolution of the EUV wave and Moreton wave, as obtained from the AIA 211 \AA, 193 \AA, 171 \AA, 94 \AA, 131 \AA, 304 \AA, and 335 \AA, SUTRI 465 \AA\ as well as SMART \ha. The deceleration and initial speed of the wave obtained by fitting the spines in the time-distance stack plots are listed in corresponding panels, where the red dotted curved lines are the fitting results using a second-order polynomial function. Panel (g) marks the oscillations of filaments.
\label{fig:tdp}}
\end{figure}

We examined temporal features of the EUV and Moreton wave by using the time-distance stack plots following a sector marked with S1 in \nfig{fig:overview} (a), which are drawn with a great circle of the solar surface from the flare kernel. Here, we only show the evolution process of the wave in the direction, i.e., S1, towards the mini filaments on northwestern limb (see \nfig{fig:overview} (a), where the filament shapes obtained from the CHASE \ha observation are overlaid on it). \nfig{fig:tdp} is the time-distance stack plots reconstructed follow S1 using running difference images (except for panel (f), which uses AIA 335 \AA\ raw images because its wave signal appears more prominently). Obviously, we can observe an inclined bright ridge in each panel, representing the wave signals propagating \textcolor{black}{from the chromosphere through the transition layer to the corona.} The wave is very bright and sharp from 01:14 UT to 01:22 UT, which is almost the same time range as the \m wave. The slope of these ridges corresponds to the speed of the wavefront. By performing a second-order polynomial fitting, we obtained the propagation speed and deceleration of the wavefront at various atmospheric heights. \textcolor{black}{As shown in \nfig{fig:tdp}, we can see that the wave propagating speeds observed in 211, 193, 171, 94, 131 and 335\,\AA\ (with a average value of about \speed{940}), which are significantly higher than the speed observed in 304\,\AA\, which was about \speed{720} (see \nfig{fig:tdp} (g)). As AIA 304 \AA\ channel primarily encompasses the He II 303.8 \AA\ line, dominate at chromosphere temperatures of log(T/K)$\sim$4.7. Thus, we believe that the signal observed in the 304 \AA\ band represents the signatures of the wave propagation in chromosphere. This can be verified from the observations in the SMART H$\alpha$ band. As shown in \nfig{fig:tdp} (i), we can see that the wave propagation speed in chromosphere observed in SMART H$\alpha$ was about \speed{720}, which is consistent with that observed in AIA 304 \AA. Furthermore, we find that the wave propagation speed in the transition region was about \speed{770} observed in SUTRI 465 \AA\ band (see \nfig{fig:tdp} (h)). This speed is higher than that measured in the chromosphere (\speed{720}) but lower than the corona value (\speed{940}).} This difference in speed further reflects that the wave front has dome-like shape: as the height increases, the projected speed become greater, a conclusion has been verified in other literature \citep{2012ApJ...753...52L}.

\subsection{The horizontal oscillations of the filaments}

As the wave propagating about \textcolor{black}{400\,Mm} arrived near the filaments on the west limb, it triggered the filaments to begin oscillating (see \nfig{fig:tdp} (g)). We selected a series of straight slice to investigate the oscillation of the filaments in the plane-of-sky (POS) direction, (i.e., the transverse horizontal oscillation). The left panels of \nfig{fig:hor_osc} is a close up region of the mini filaments in the AIA 171\,\AA, CHASE \ha, and SMART \ha, where three mini filaments, F1, F2 and F3, can clearly be identified. We draw three sets of straight lines (\nfig{fig:hor_osc} (a1)-(a3)), overlaying on the corresponding filament, to study their oscillation in the directions perpendicular to their axes. The starting of lines for F1 and F3 are located at the center of the circle. Since the axis of F2 is difficult to distinguish, we took its centroid as the center and draw 12 diameters at 15$^\circ$ intervals to analysis its oscillation in the POS. The starting points of each diameter is marked with a solid dot at its end. 
From the time-distance stack plots shown in \nfig{fig:hor_osc} (b1)$-$(b3) derived from raw AIA 171 \AA\ images, we found that filaments F1 and F2 exhibited significant oscillations. These oscillations commenced around 01:22 UT, and lasted for 5 and 3 cycles, respectively. In contrast, filament F3 remained stationary during this period. To obtain the oscillation parameters, we fit the oscillation profile with a damping function in the form of $f(t)=Ae^{(-t/\tau)}cos(2\pi/T+\phi)$, where $A$, $T$, $\tau$, and $\phi$ are the initial amplitude, period, damping time, and initial phase, respectively. The fitting results reveal that filament F2 exhibits an oscillation period of 1108\,s and a amplitude of 2.5\,Mm, both of which are significantly larger than those of filament F1 (520\,s and 1.5\,Mm, respectively). All time-distance stack plots results can be found in \nfig{fig:a_f13} and \nfig{fig:a_f2} of the Appendix. We also employ this fitting method to derive the vertical oscillation parameters (see \nfig{fig:ver_osc1}).

\subsection{The vertical oscillations of the filaments}

\begin{figure}
	\centering
	
	\includegraphics[width=.75\linewidth]{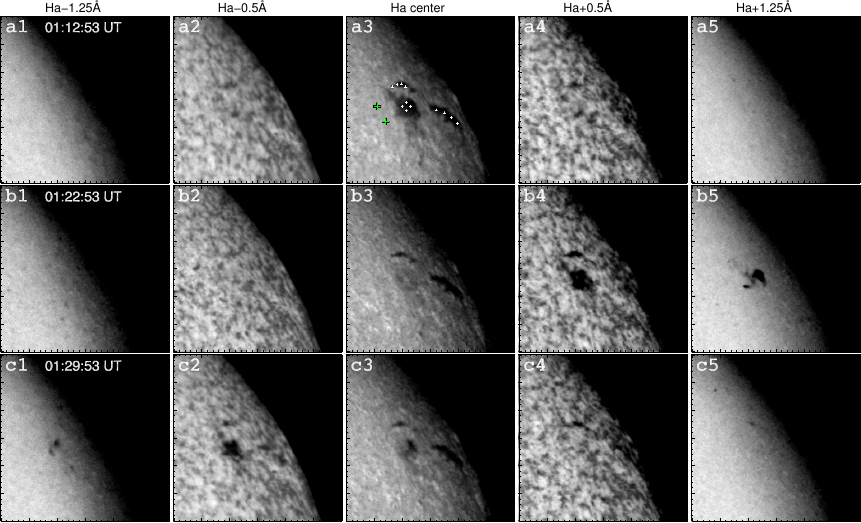}
	\caption{ Time sequence of SMART \ha images show the oscillation of the filaments in the LOS direction at three different time. The five columns, from left to right, represent images from bands -1.25 \AA, -0.5 \AA, 0.0 \AA, +0.5 \AA\ and +1.25\AA\ images shifting from the \ha line center (6562.8\,\AA ),  respectively. Images in the same row are taken at a near simultaneous time refer with the \ha center image. The four white marked points on F1, F2 and F3 in panel (a3) are the positions where Doppler velocities are measured, while the two green points are used to get the background intensity.
		\label{fig:winking}}
\end{figure}

\nfig{fig:winking} presents the time sequence of SMART \ha line-center and line-wind images to illustrate the motion of the filaments in the LOS direction at three different stages. The middle column shows the \ha line-center images, while the left (right) column display the images in the blue (red) wings ($\pm1.25$\,\AA\ and $\pm0.5$\,\AA), respectively. The first row presents the state of the filaments before the oscillation begins, and these three filaments can be clearly seen in the \ha center image (see panel (a3)). Around 01:22 UT, F1 and F2 appear in the red wing images (see \nfig{fig:winking} panels (b3)$-$(b5)), suggesting that these two filaments undergo a downward motion. Meanwhile, the intensity of F1 and F2 decreases in the \ha center image. As the wavelength increases, the intensity of the filaments decreases. In the \ha $+$1.25\,\AA\ image, F2 can still be seen while F1 disappears. Approximately seven minutes later, F1 and F2 appear in the blue wing images (see \nfig{fig:winking} panels (c1)$-$(c3)), indicating that they start to move upward. F1 is visible in the image \ha\ $-$0.5\,\AA\, but its intensity is lower than that of F2. In the \ha\ $-$1.25\,\AA\ image, F1 disappears completely while F2 is still partially visible. Compared to filament F1 and F2, the winking signal of filament F3 is relatively less obvious: only some faint signals can be find in red wing (panel (b4)) and blue wing ( panel (c2)).

\begin{figure}
	\centering
	\includegraphics[width=.75\linewidth]{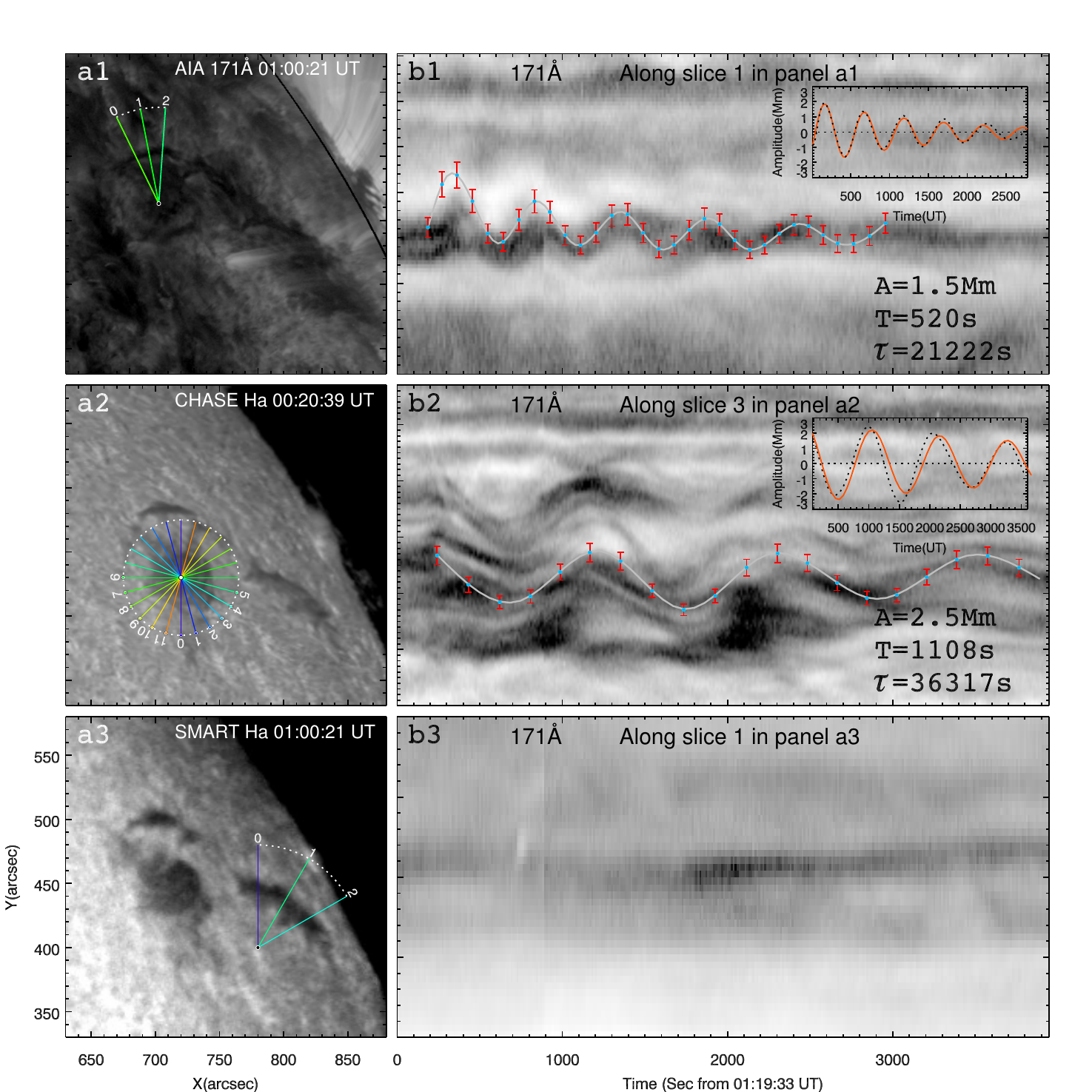}
	\caption{The horizontal oscillation of the filaments. Panels (a1)-(a3) show the filaments in AIA 171 \AA, CHASE \ha and SMART \ha images, where the color lines marked the path to get time-distance stack plots to study their oscillation in different directions. In panel (a1) and (a3), the paths start at the center of the circle, while in panel (a2), the paths separated by 15$^\circ$ angle are located on the arc and are marked with solid circular dots.  Panel (b1) shows the horizontal oscillation of filament F1 along the slice 1 in panel (a1). Panel (b2) is an example of the filament oscillation in the direction of slice 3 in panel (a2), while  panel (b3) is an example of filament F3's oscillation in direction of slice 1 of panel (a3). The gray curves in panel (b1) and (b2) are the oscillation curve that manually marked the positions using for fitting. The black curve is the detrended of curve of the oscillation curve obtained by subtracting the smooth velocity using a 10 minutes boxcar, where the red curve is its fitting result using the formula $f(t)=Ae^{(-t/\tau)}cos(2\pi/T+\phi)$. The fitting results are listed in corresponding panel.
\label{fig:hor_osc}} 
\end{figure}
\begin{figure}
	\centering
	\includegraphics[width=.75\linewidth]{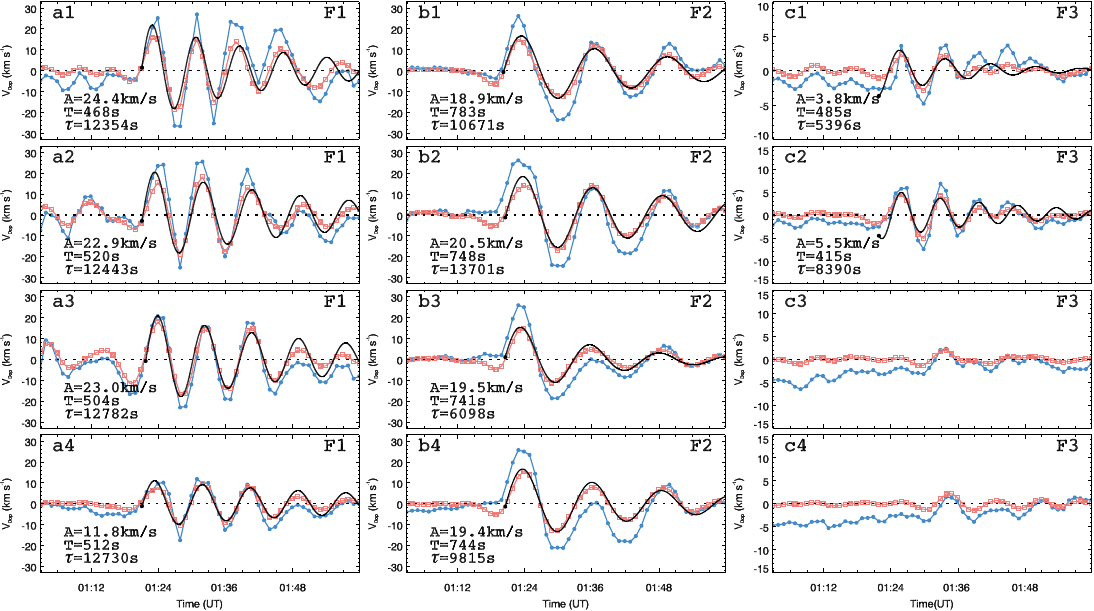}
	\caption{Measurement of the Doppler velocity of the oscillation filaments. The Doppler velocities obtained from sampling points on filaments F1, F2 and F3 are shown in (a1)-(a4), (b1)-(b4) and (c1)-(c4), respectively. In each panel, the blue curve with circle dot is the derived Doppler velocity, while the red curve with diamond is the detrended velocity. The back curve is a fit to the detrended velocity using the formula $f(t)=Ae^{(-t/\tau)}cos(2\pi/T+\phi)$. The fitted velocity amplitude (A), oscillation period (T) and damping time ($\tau$) are listed in corresponding panel.
		\label{fig:ver_osc1}}
\end{figure}

To obtain the detailed parameters of vertical oscillation of these filaments, we select four sample points on F1$-$F3 respectively (marked by white cross points in \nfig{fig:winking} panel (a3)) for Doppler velocity. To obtain the Doppler velocity of the filament, we first measure the intensity contrast profile of a sample point on the filaments by dividing the nearby average background intensity (marked with green cross in \nfig{fig:winking} panel (a3)). Subsequently, we fit the derived intensity contrast profile using the \texttt{cloud model} \citep{1964PhDT........83B} to determine the corresponding Doppler velocity of the sample points. The detailed analysis results of the three filaments' oscillations along the LOS is shown in \nfig{fig:ver_osc1}. The blue curves in each panel are the Doppler velocities acquired from four sampling points of filaments F1, F2 and F3, respectively. These periodic fluctuations observed in the velocity measurements demonstrate the oscillatory motion of the filaments in the LOS direction. Additionally, it is evident that both filaments, F1 (panels (a1)$-$(a4)) and F2 (panels (b1)$-$(b4)), initially exhibit a stronger redshift signal when the wave approaches them around 01:22 UT. Afterward, blueshift and redshift alternate. This evolutionary pattern suggests that the wave first pushed the filaments downward due to a strong compression originating from its dome-shaped propagating wavefront. Following the passage of the pulse, the filaments begin to oscillate with their own characteristic period. By fitting the detrended Doppler velocities using the same method mentioned in \nfig{fig:hor_osc}, we find that: the period for F2, approximately 740$-$780\,s, is significantly longer than that of F1, which is around 470$-$520\,s. In contrast, the velocity amplitude for F1, ranging from 12 to \speed{24}, is slightly higher than that of F2, which ranges from 19 to \speed{21}.  Notably, the four sample points of F2 were relatively concentrated in the central region of filament F2, resulting in approximately identical oscillation start times observed in the Doppler velocity map. However, the four sampling points of F1 were distributed along the filament axis from left to right, leading to a continuous variation in the oscillation onset times. For filament F3, its left portion exhibited significant oscillations with amplitudes of 3.8 to \speed{5.5} and periods of approximately 415$-$485\,s (see \nfig{fig:ver_osc1} (c1) and (c2)). In contrast, its right portion showed no significant oscillation. Since the oscillation onset at the two left sampling points (01:22 UT) coincided with the wave's arrival and their amplitudes were considerably larger than those at the two right points, we consider the left two points to accurately reflect the oscillation of filament F3, despite minor oscillatory signals also being present at the right two points.

\section{Discussion and conclusions} 
\label{se:discussion}
In this paper, we report a dome-shaped wave that was observed simultaneously in different layers of the solar atmosphere, which triggers oscillations in three filaments with varying amplitudes: two exhibit large-scale oscillations, while one shows small-scale oscillations. In the initial stage, the EUV wave observed in AIA/SDO EUV wavebands, the surface part of the wave, was co-spatial with the \m wave observed in the chromosphere using the SMART \ha images. We also detected the wave signals in the transition region between the chromosphere and corona using SUTRI 465 \AA\ images, which have often been overlooked in previous studies. During the free propagation stage, the speed of the wave increases with the height: it is about \speed{720} in the chromosphere, \speed{770} in the transition region, and \speed{940} in the corona. This difference in speed indicates that the wave has a dome-shaped. After propagating about \textcolor{black}{400\,Mm}, it interacted with three filaments along its propagation path, causing the first two to exhibit oscillation simultaneously in both horizontal and vertical directions, while the third one only oscillated in the vertical direction. Interestingly, both F1 and F2 were large-scale oscillations with a velocity amplitude about \speed{20}, while F3 was just oscillated on a small scale with an amplitude about \speed{4}. The vertical oscillations of filaments F1, F2 and F3 result in the winking behavior in the \ha observations. The oscillation periods in horizontal and vertical of filament F1 are both approximately 500\,s. In contrast, for filament F2, the periods show a significant difference, with values of 1100\,s and 750\,s, respectively.
Filament F3 only exhibits oscillation in the vertical direction with a period of about 450\,s.

 \m waves in the chromosphere and EUV wave in the corona are intriguing large-scale phenomena in the solar atmosphere. Since the \alf speed in the chromosphere is approximately \speed{100}, the \m waves in the chromosphere cannot propagate to a distance of about 100\,Mm at a speed of approximately \speed{1000}. Instead, the waves will rapidly dissipate because of the excessively high Mach number. \cite{1968SoPh....4...30U} proposed that the pressure pulse from a flare generated a dome-shaped fast-mode wave or shock wave in the corona, propagating outward at a speed of approximately \speed{1000}. The base of the wave sweeps across the chromosphere, generating a \m wave. In other words, the \m wave is the imprint of the base of the fast-mode wave or shock wave in the corona on the chromosphere. The coronal wave, i.e. EUV wave, has been extensively studied in the past twenty years. Their physical nature and driven mechanism have finally been unified, namely, they are a fast-mode magnetosonic wave driven by the lateral expansion of the accompanying CMEs \citep{2002ApJ...572L..99C,2011ApJ...738..160M,2012ApJ...746...13L,2020ApJ...889..139M,2022ApJ...928...98H,chen2022}. Although the model of \cite{1968SoPh....4...30U} is widely accepted, observations indicate that most EUV waves are not accompanied by \m waves. Regarding this issue of \m waves being rarely observed, \cite{2023ApJ...949L...8Z} found that EUV waves and \m waves exist only in eruptions that are sufficiently inclined. The AIA 304 \AA\  images show that the ejection of the filament reported here erupted obliquely towards the northwest, leading to the generation of the \m wave propagating in the same direction. This is consistent with the result proposed by \cite{2023ApJ...949L...8Z}. \cite{2022ApJ...933..148C,2024ApJ...977L..26C} conducted numerical experiments to investigate the cause of the inclined eruption of the filament, discovering that the occurrence of a catastrophe in an asymmetric and non-force-free environment is the primary reason for this phenomenon. We anticipate future numerical simulation of non-radial eruptions leading to the \m waves. This would enable a more detailed study of the relationship between the \m wave and the erupting filaments.

According to the theoretical interpretation in \cite{1968SoPh....4...30U}, the shock wave has a dome-shaped structure and appears as a circular shape when projected onto the solar disk. When this shock wave intersects with the chromosphere, it will lead to the formation of the \m\ wave. This raises the question: since the transition between the corona and chromosphere is inevitable for the propagating shock, one would expect to observe wave signals in this region as well. However, in reality, little attention has been paid to such wave signals in the transition zone. 
 The wave, reported here, was simultaneously observed in the AIA EUV wavebands, SUTRI 465 \AA\ and SMART H$\alpha$ images, covering a broad temperature range from the low corona, transition region, to the chromosphere, indicating that the EUV wave and its chromospheric counterpart, namely, \m wave, as well as the response in the transition are observed. Notably, we find that a decreasing wave speed in the initial phase, moving from the lower corona towards the chromosphere: about \speed{940} in the corona, \speed{770} in the transition region, and \speed{720} in the chromosphere, This is probably a result of the dome-shaped wavefront structure, causing a delay in the disturbance reaching the lower atmosphere. The expanding dome-shaped of the EUV always gets the evidence from the limb observation in previous observation \citep{2011ApJ...738..160M,2012ApJ...745L...5C,2012ApJ...753...52L,2021SoPh..296..169Z}. In this study, we present the first empirical verification of the wave front's geometry achieved by analyzing the velocity differences at different solar atmospheric heights.

Large-amplitude oscillations (greater than \speed{20}) are the periodic motions of the entire filament (or most of it) body often associated with \m waves or EUV waves \citep[e.g.,][]{2002PASJ...54..481E,2008ApJ...685..629G,2012ApJ...761..103G,2012ApJ...745L..18A,2013ApJ...773..166L,2023ApJ...959...71D}. The earliest reported observations of filament oscillation can be traced back to the 1930s \citep{1935MNRAS..95..650N}. Although horizontal and vertical oscillation of filaments has been widely studied, the simultaneous horizontal oscillation and the winking phenomenon in a filament has rarely been reported \citep{2018ApJS..236...35L}. To the best of our knowledge, only a few of papers have reported the simultaneous presence of similar horizontal and vertical oscillation in a filament: \cite{2006A&A...449L..17I}, \cite{2023ApJ...959...71D} and \cite{2025PhFl...37c7177O}. Since the filament was located far from the disk center in the southern solar region, the horizontal oscillation reported by \cite{2006A&A...449L..17I} is subject to considerable uncertainty due to foreshortening effects, as already noted by the original authors. Moreover, given the relatively low resolution of available observational data, it is quite difficult to accurately determine filament position changes based solely on two images. In contrast, using SMART \ha line-wing observations, \cite{2014ApJ...786..151S} detected successive episodes of filament winking. Interestingly, while these filaments exhibited clear vertical oscillations in the SMART data, they showed neither transverse horizontal nor longitudinal oscillations in the simultaneous high-resolution observations from AIA/SDO. \cite{2023ApJ...959...71D} and \cite{2025PhFl...37c7177O} provided evidence for simultaneous horizontal and vertical oscillations. However, along the LOS direction, they only presented velocity parameters without explicit information regarding oscillations.

A significant application of filament oscillations and coronal seismology is to calculate parameters such as the ambient magnetic field that cannot be directly measured \citep[e.g.,][]{2014ApJ...786..151S,2018ApJS..236...35L,2023ApJ...959...71D,2020A&A...638A..32Z,2023A&A...675A.169L}. Only based on the correct analysis of the oscillation modes can we calculate the corresponding magnetic field using relevant parameters. By using the formula for vertical oscillation of filament given in \cite{1966ZA.....63...78H}：\[B_r^2 = \pi \rho r_0^2 \left[ 4\pi^2 \left(\frac{1}{T}\right)^2 + \left(\frac{1}{ \tau }\right)^2 \right]\]
We can obtain the radial component of the magnetic field \(B_r\), where $\rho=N_Hm_H$ ($N_H$ and $m_H$ are the average number density and mass of hydrogen atom respectively) is the density of the filament material, \(r_0\) is the scale height of the filament, \textit{T} is the vertical oscillation period of the filament, and $\tau$ is the decay time. We can substitute $m_H=1.67\times10^{-24}$ $\text{g}$, $N_H=3\times10^{10}$ $\text{cm}^{-3}$ \citep{2010SSRv..151..243L}, \(r_0\)=$3\times10^9$ $\text{cm}$ \citep{1966ZA.....63...78H}, and the measurement data from this event into the formula for calculation. According to the fact that the vertical oscillation average periods of F1, F2 and F3 are 501\,s, 754\,s, 450\,s; the mean decay time are 12577\,s, 10071\,s and 6893\,s, we can calculate that the value of \(B_r\) at the position of F1, F2 and F3 is 14.9\,G, 9.9\,G and 16.6\,G, 
respectively. This value falls within the range of 2$-$30 G provided by \cite{1966ZA.....63...78H} and is also fairly consistent with previous research results \citep{2014ApJ...786..151S,2017ApJ...850..143L,2023ApJ...959...71D}. According to the period equation for horizontal oscillation given by \cite{1969SoPh....6...72K}:\[P=4\pi L B_l^{-1}\sqrt{\pi\rho}\]
Among them, \textit{P} refers to the period of horizontal oscillation, \textit{2L} is the length of the filament, $\rho$, as mentioned above, is the density of the filament. By substituting the measured period of 525\,s (1115\,s) for the F1 (F2) horizontal oscillation and the estimated length 60.19\,Mm (83.10\,Mm) (detailed measurement methods in the Appendix), 
we can calculate the magnetic field strength at the position of F1 (F2) along axial direction of the filament to be 28.6\,G (18.6\,G).

The interaction mechanism between solar filaments and waves is an important and complex issue. \cite{2014ApJ...795..130S} proposed a simple model where the 
oscillation mode (transverse or longitudinal) depends on the relative orientation of the wave normal vector and the filament axis: transverse oscillations occur if the normal vector is perpendicular to the axis, while longitudinal oscillations occur if it is parallel. However, this simple picture is challenged by observations. For example, \cite{2014ApJ...786..151S} reported a case where the filament F1 was parallel to the wave normal vector, filaments F2, F3, and F4 were perpendicular to the wave normal vector, yet only vertical oscillations were observed, with neither longitudinal nor horizontal oscillations detected. Similarly, \cite{2018ApJ...860..113Z} found only vertical oscillations in a filament where the wave normal vector was perpendicular to the filament axis. These contradictory observations demonstrate that the wave-filament interaction is still a complex issue and requires further investigation, particularly through numerical simulations and multi-wavelength, multi-viewpoint observations. 
To the best of my knowledge, only a few numerical simulations have focused on this complete process, from wave generation and propagation to its interaction with distant filaments, such as \cite{2020A&A...637A..75L} and \cite{2025A&A...696A.158L}. We anticipate that further 3D simulations of the wave-filament interaction will offer more detialed insights into their interaction process.

In this study, our simultaneous AIA and SMART observations of filament F1 reveal a similar oscillation period of approximately 500\,s in both the apparent horizontal and vertical directions. Furthermore, both apparent oscillations exhibited 5 cycles with nearly identical onset times, around 01:22 UT. Based on these similarities, we make the bold conjecture that filament F1 might, in fact, undergo only genuine vertical oscillations relative to the solar surface. The apparent horizontal oscillation could then be a projection effect. This conjecture is supported by previous findings, such as those by \cite{2014ApJ...786..151S} and \cite{2018ApJ...860..113Z}, who reported observing exclusively vertical oscillations in filaments located near the disk center and at the limb, respectively. These results lend credence to the idea that the apparent horizontal oscillation of filament F1 reported here might be solely due to the projection of its underlying vertical motion. This highlights a key challenge: the same intrinsic oscillation mode can manifest as different observed behaviors depending on the filament's location and orientation relative to the observer. Consequently, accurately determining the true oscillation mode and mechanism is particularly difficult when observing from a single viewpoint and in a single spectral band. 

In contrast, filament F2 exhibited a significant difference in oscillation periods between the apparent horizontal and vertical directions. This difference strongly suggests that F2 underwent genuine, distinct horizontal and vertical oscillations, rather than a single projected motion. As confirmed in Appendix \nfig{fig:a_f2}, analysis of 12 slices around F2 revealed oscillation signals in multiple directions. The most prominent signals were observed in slices 6, 7, and 8, indicating that horizontal oscillation was the dominant component among the observed motions. Regarding filament F3, it exhibited a small-scale oscillation in vertical direction with a velocity amplitude about 3.8$-$\speed{5.5} in its left portion, which is consist with previous reports indicating that the small-scale oscillations have local characteristics \citep{1986SoPh..104..313T,1993A&A...277..635B,2002A&A...393..637T,2008ApJ...676L..89B}. Generally, small-amplitude oscillations are not related to an external trigger \citep{2018LRSP...15....3A}. However, we propose that the small-amplitude oscillation of filament F3 reported here was triggered by the \m wave as its onset oscillation coincided with the arrival of the \m wave. 


In conclusion, the excellent spatial and temporal resolution of the observations presented here provides a better understanding of the physical properties of filaments, EUV waves, and \m waves. These insights offer potentially valuable inputs and constraints for theoretical models. Furthermore, as the Sun progresses through solar cycle 25 and approaches its maximum activity \citep{2023SCPMA..6629631C}, it presents a prime opportunity to study the interactions between \m waves and filaments.

\appendix
\label{se:appendix}
\section{The horizontal oscillation of the filaments}
\begin{figure}
	\centering
	
	\includegraphics[width=.75\linewidth]{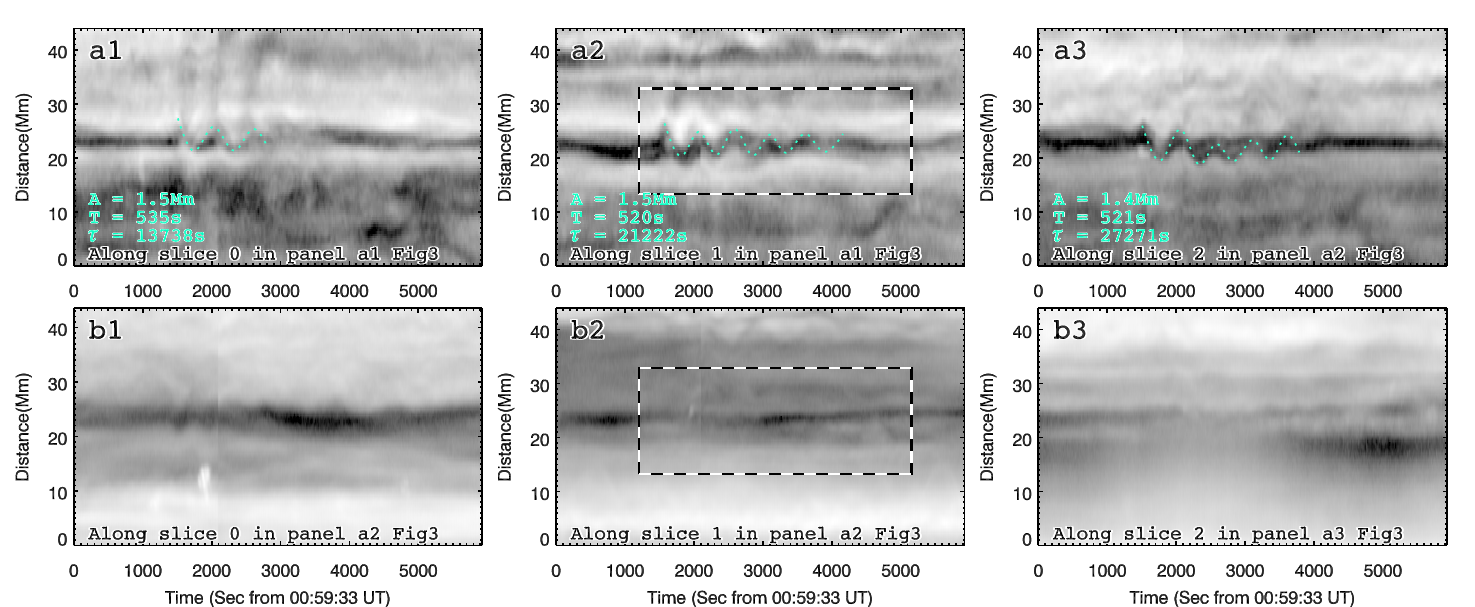}
	\caption{The time-distance stack plots for filaments F1 and F3 along the slices marked 0, 1 and 2 in panels a1 and a3 of \nfig{fig:hor_osc}, using the AIA 171\,\AA\ raw images. The green curves in top row are the oscillation curves that manually marked the positions using for fitting, and the fitting results are listed in corresponding panels. The boxes are the FOV of panel b1 and b3 in \nfig{fig:hor_osc}. 
		\label{fig:a_f13}}
\end{figure}

\begin{figure}
	\centering
	
	\includegraphics[width=.75\linewidth]{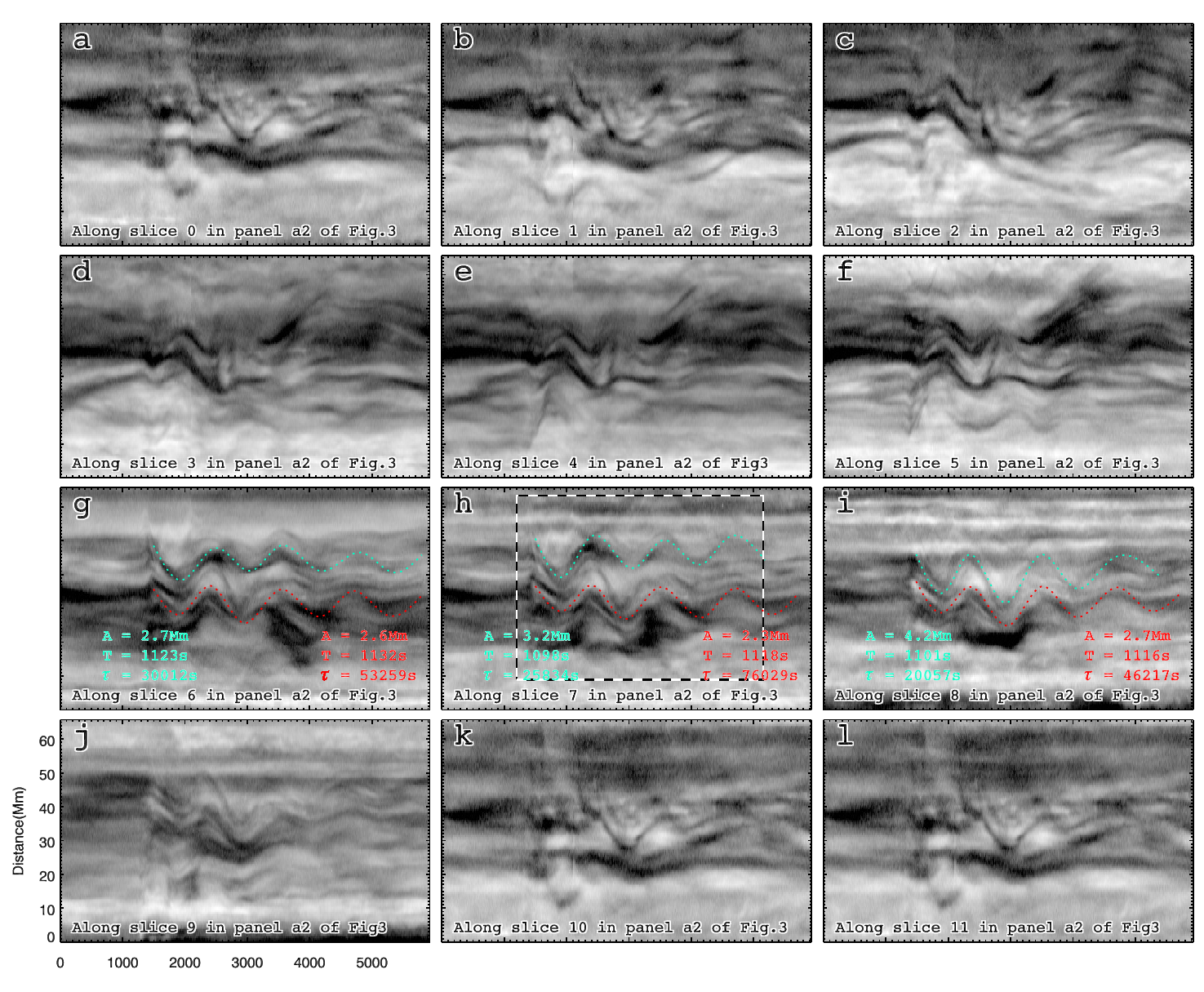}
	\caption{Time-distance stack plots are generated for 12 slices in \nfig{fig:hor_osc} a2, with each slice spaced at 15$^\circ$ intervals and centered on the filament mass F2, using 171 \AA\ raw images. Third row is the oscillation of filament F2 in the directions slice 6, 7 and 8 in \nfig{fig:hor_osc} panel a2, where the green and red curves are the oscillation curve that manully marked the positions using fore fitting and their corresponding fitting results are listed with corresponding colors in each panel. The box in panel S7 is the Fov for \nfig{fig:hor_osc} b2. 
        \label{fig:a_f2}}
\end{figure}

As show in \nfig{fig:hor_osc}, We selected three sets of paths for space-time analysis. Since the spines of filament F1 and F3 were easy to be identified, we select 3 paths perpendicular to the filament axis for filament F1 and F3 respectively to detected they oscillation in horizontal directions. However, it is difficult to judge whether the filament F2 oscillation along axis direction or perpendicular to it. Thus,  we placed 12 slices centered on the filament mass F2 in different directions, where the 12 directions are uniformly distributed in the azimuthal directions as shown by the color solid lines in panel a2 of \nfig{fig:hor_osc}, using the method proposed by \cite{2017Ap&SS.362..165C} and further developed by \cite{2023MNRAS.520.3080T}.The corresponding time-distance stack plots of 171 \AA\ images along these slices are shown in \nfig{fig:a_f13} and \nfig{fig:a_f2}. The starting points of these paths are all marked with solid small dots located at one of their endpoints.

As illustrated in \nfig{fig:a_f13}, filament F1 exhibits significant oscillations in all three directions, while filament F3 remains stable during F1's motion. From the fitting results, we can find that the oscillation periods in three directions are respectively 530\,s, 520\,s and 521\,s, with a average value of 524\,s. Obviously, This periods is consistent with it vertical oscillation that obtained from the Doppler images. From \nfig{fig:a_f2}, one can find that the oscillation pattern is the most significant along the slice 7, which is almost perpendicular to the local filament spine. That fact that the direction of the most significant oscillation is perpendicular to the filament spine implies that the filament oscillation is a horizontal oscillation.

\section{measurement of the filaments' length}
In actual observations, a solar filament should have a real 3D structure with variations in height. To comprehensively analyze the 3D length of a solar filament, stereoscopic observations from at least two different perspectives are required \citep{2023MNRAS.520.3080T}. However, currently, only single perspective observations from the Earth are available. Therefore, we assume that the solar filament has no height variations, that is, the solar filament is distributed on a spherical surface with a fixed radius (the solar surface). Assuming that filaments F1 and F2 are uniform cylinders, we select 10 points uniformly along their axes from one endpoint to another respectively (in helioprojective cartesian (HPC) coordinates). Using the functions in the \texttt{astropy} and \texttt{sunpy} libraries in Python, we convert the HPC coordinates of these ten points into heliographic Stonyhurst (HGS) coordinates. Keeping the longitude and latitude unchanged, the radius is adjusted to the solar radius $R_{\odot}$. We apply the formula for converting spherical coordinates to Cartesian coordinates to transform the new HGS coordinates into 3D Cartesian coordinates (output the value of x, y, z in units of the solar radius $R_{\odot}$, and finally take $R_{\odot}$ as 696.34 Mm). Lastly, we calculate the corresponding distances using the x, y, z values of every two adjacent points. Summing up the spacings of the 10 sample points yields the estimated length of the filament. This approximation method can yield the most optimal estimation results under the current observational condition. Although it will generate certain errors, these errors are acceptable because the variation in the height of the solar filament is not significant.

\section{Acknowledgments}
We would like to thank Prof. \texttt{Pengfei Chen} in Nanjing University, Dr. \texttt{Song Tan} in Leibniz-Institut f\"{u}r Astrophysik Potsdam (AIP) and Dr. \texttt{Qingmin Zhang} in Purple Mountain Observatory of the Chinese Academy of Sciences for valuable discussions and suggestions. This paper utilized data from SDO, SMART, SUTRI and CHASE. This work is supported by the Natural Science Foundation of China (12303062,12373063,12103016), the Sichuan Science and Technology Program (2023NSFSC1351, 2025ZNSFSC0315),  the Project Supported by the Specialized Research Fund for State Key Laboratories, Key Laboratory of Detection and Application of Space Effect in Southwest Sichuan at Leshan Normal University, Education Department of Sichuan Province(No.ZDXM20241002), and the Fund of Shenzhen Institute of Information Technology (SZIIT2025KJ003, HX-0951). We also acknowledge Sichuan Normal University Astrophysical Laboratory Supercomputer for providing the computational resources.

\vspace{5mm}
\end{CJK*}
\end{document}